\documentclass[prb,  onecolumn, square,showpacs, aps] {revtex4}
\usepackage[dvips]{graphicx}

\begin{document}

\title{Effect of the metal-to-wire ratio on the high-frequency magnetoimpedance 
of glass-coated CoFeBSi amorphous microwires}

\author{R. Valenzuela\P, A. Fessant, J. Gieraltowski and C. Tannous} 
 \affiliation{Laboratoire de Magn\'{e}tisme de Bretagne, CNRS-FRE 2697, 
Universit\'{e} de Bretagne Occidentale, BP 809, Brest CEDEX, 29285, France \\
\P Departamento de Materiales Met\'{a}licos y Cer\'{a}micos,
Instituto de Investigaciones en Materiales, Universidad Nacional 
Aut\'{o}noma de M\'{e}xico, P.O. Box 70-360, Coyoac\'{a}n, Mexico D.F., 04510, Mexico}

\begin{abstract}
High frequency [1-500 MHz] measurements of the magnetoimpedance (MI) of 
glass-coated Co$_{69.4}$Fe$_{3.7}$B$_{15.9}$Si$_{11}$ microwires are carried 
out with various metal-to-wire diameter ratios. A twin-peak, anhysteretic 
behaviour is observed as a function of magnetic field. A maximum in the 
normalized impedance, $\Delta Z$/$Z$, appears at different values of the 
frequency $f$, 125, 140 and 85 MHz with the corresponding diameter ratio $p$ = 
0.80, 0.55 and 0.32. We describe the measurement technique and interpret our 
results with a thermodynamic model that leads to a clearer view of the 
effects of $p$ on the maximum value of MI and the anisotropy field. The 
behavior of the real and imaginary components of impedance is also 
investigated; they display a resonance that becomes a function of the DC 
field $H_{DC}$ for values larger or equal to $H_{K}$ the circumferential 
anisotropy field for each $p$ value. These results are interpreted in terms
 of a rotation model of the outer shell magnetization.
\end{abstract}


\maketitle

Magnetostriction.

\section{INTRODUCTION}

When a ferromagnetic conductor is traversed by a current of low amplitude 
and high frequency, its impedance, or rather its Magnetoimpedance (MI) can 
be altered by applying a DC magnetic field. This phenomenon, first described 
[1] in the 1930's, has been receiving special attention over the last 15 
years [2,3] due to its potential technological applications [4,5] in 
sensors, devices and instruments. Its fundamental physics is also being 
deeply examined [6]. MI has been observed in a wide variety of materials, 
geometries and structures, particularly in amorphous wires having 
diameters of a few hundred microns. Wires with smaller diameters (a few 
microns) coated with a glass sheath show an increase of the working 
frequency, and introduce an additional structural feature that alter the 
physical parameters [7,8]. 

Since glass exerts some mechanical stress on the 
metallic wire, a change in the magnetic response is expected; other works 
have rather investigated the effect of external stress [9,10], as well as 
that of various annealing methods [11,12]. Consequently, it is of interest 
to finely tune the physical properties through the control of the thickness 
and nature of the glass sheath. In this paper, we report on MI measurements 
of Co-rich amorphous microwires with various ratios of the metal-to-wire 
diameter, in the [1-500 MHz] frequency range, carried out with a novel [13] 
broadband technique. This technique allows a complete determination of MI as 
a function of both frequency and magnetic field. 

In contrast with most 
published works, where MI is measured at a single frequency, or at a 
discrete set of frequencies, this technique provides a quasi-continuous 
ensemble of results over the 1-500 MHz frequency range. The effects of the 
thickness of the glass sheath are clearly illustrated and the variation of 
the anisotropy field $H_{K}$ is evaluated directly as a function of stress. 

Additionally, we carried out an analysis of the real and imaginary 
components of impedance. The resonance character observed can be attributed 
to the rotation of the outer shell magnetization as described further below. 
Possible sensor applications are discussed in the conclusion.

\section{MEASUREMENT TECHNIQUES}

Glass-coated amorphous microwires of nominal composition 
Co$_{69.4}$Fe$_{3.7}$B$_{15.9}$Si$_{11}$ were prepared by fast cooling with 
the Taylor-Ulitovski technique \cite{Torcunov}. Several metal-to-wire ratio values, 
$p =\phi_{m}/\phi_{w}$, with $\phi_{m}$ the metallic core diameter and $\phi_{w}$ the total 
wire diameter, were produced and characterized. For values of metal core 
diameters of 24, 12 and 7 $\mu $m, with corresponding total diameters of 30, 
21.8 and 21.9 $\mu $m, we get the ratio $p$ = 0.80, 0.55 and 0.32, 
respectively. In order to make electrical contacts, the glass sheath was 
etched away over a few mm on both microwire ends, with a solution of 
hydrofluoric acid. Silver paste contacts were then made in order to proceed 
with the electrical measurements. 

MI measurements were carried out in the 
[1-500 MHz] range, on pieces of microwires $\sim $12 mm in length, with an 
HP 8753C Network Analyzer using a novel broadband measurement technique 
described in [13]. Helmholtz coils served as source of axial DC magnetic 
fields ranging from -80 Oe to +80 Oe. Ferromagnetic resonance 
(FMR) measurements were carried out 
using 3 mm long samples, in a JEOL JES-RES 3X spectrometer operating at 9.4 
GHz (X-band), at room temperature. Non-resonant low-field absorption (LFA) 
measurements were taken using a JEOL ES-ZCS2 zero-cross sweep unit, which 
compensates for any magnetic remanence, allowing precise determination of 
low magnetic fields around zero.

\section{EXPERIMENTAL RESULTS AND MODELING}

The results obtained are plotted in a continuous 3D representation 
of $\Delta Z/Z$, with:
\begin{equation} 
\Delta Z/Z=\frac{[Z(H_{DC})-Z(H_{DC}=80 \mbox{ Oe})]}{Z(H_{DC}=80 \mbox{ Oe})}, 
\end{equation}
where $Z(H_{DC})$ is the total impedance modulus
$Z =\sqrt{ (Z'^{2} + Z''^{2})}$, with $Z$' the real part and $Z$'' the imaginary part of impedance. 
$Z(H_{DC})$ is a function of the DC field, $H_{DC}$ (in Oersteds), and 
frequency, $f$. The results for $p$ = 0.8 are shown in Fig. 1. A symmetrical 
double peak MI plot is obtained as a function of $H_{DC}$; the peaks are 
associated with $\pm  H_{K}$, the circumferential anisotropy field. We 
obtain $H_{K}  \sim $ 3.5 Oe and no hysteresis by cycling the DC field 
$H_{DC}$.

Regarding frequency $f$, the MI shows a maximum of $\sim $ 250 {\%} at 
about 100 MHz. Similar plots were obtained with the other $p$ ratios, albeit 
with significant differences in the values of the anisotropy field and peak 
frequency values. Instead of making a comparison at a single frequency, as 
typically done, we choose the frequency at which the maximum in $\Delta Z/Z$ appears
 and compare results as a function of $H_{DC}$, as shown in Fig. 2. 
Note that the sample with $p$ = 0.32 exhibits, as a function of field, several 
peaks that can be associated with a distribution of the anisotropy axis 
orientation. This introduces a large uncertainty in the numerical value of 
the anisotropy field $H_{K}$. 

Figure 2 shows clear trends in the results: MI response increases as $p$ 
increases, while the anisotropy field decreases. $p$ indicates the importance 
of the metal core with respect to the total diameter of the wire and the 
stress increases as the thickness of the glass sheath increases. During 
fabrication, glass-coated microwires are subjected to strong stresses, 
generally proportional to the thickness of the glass coating that varies 
inversely proportional to $p$. The origin of such stresses can be understood, 
since glass possesses a smaller thermal contraction coefficient than metals. 
In the cooling process, the metallic core tends to contract faster and more 
substantially than the surrounding glass sheath; however glass hampers such 
contraction. 

Torcunov [14] modeled the thermoelastic and quenching stresses that occur in 
glass-coated wires and evaluated with a thermodynamic model the stress 
components in terms of their axial $\sigma_{zz}$, radial $\sigma_{rr}$ and azimuthal 
$\sigma_{\phi \phi }$ components (in a cylindrical system of coordinates $(r, \phi, z)$ 
with the $z$ direction along the wire axis). The following expressions 
(providing the Poisson's coefficients of the glass and metal are equivalent 
$\nu _{g}  \sim   \nu _{m}  \sim $ 1/3 ) are obtained and adapted to 
our case: 

\begin{eqnarray}
\sigma_{rr} & = & \frac{\epsilon E_{g} (1-p^2)}{(\frac{k}{3} +1) (1-p^2) +\frac{4 p^2}{3} } \\
 \sigma_{\phi \phi} & = &  \sigma_{rr}  \\
 \sigma_{z z} & = &  \sigma_{rr}  \frac{(k+1)(1-p^2) + 2 p^2}{k (1-p^2) + p^2} 
\end{eqnarray}

where $E_{g}$ is the glass Young modulus, $k=E_{g}/E_{m}$ and $E_{m}$ is the metallic wire
Young modulus. \\

The term $\epsilon$ is given by the difference of the glass
and metal expansion coefficients $\alpha_g, \alpha_m$ (respectively) times the difference
of the minimum glass solidification  temperature $T^{*}$ and room temperature $T$,
$\epsilon=(\alpha_m-\alpha_g)(T^{*}-T)$. \\

We apply this variation to the anisotropy field $H_K=2K_\sigma/\mu_0 M_s$ with
$K_\sigma$ the anisotropy constant of the wire 
under stress $\sigma$. We consider that the latter induces a change in the 
anisotropy constant according to 
$K_\sigma= K_{(\sigma=0)} -\frac{3}{2}\lambda_s (\sigma_{zz} - \sigma_{\phi \phi}) $
with the additional assumption of no extra
applied stress ($\mu_0$, $M_s$ and $\lambda_s$ are vacuum permeability and saturation
magnetization and magnetostriction respectively). Using physical 
parameters of wires [15] with a composition 
(Co$_{0.94}$Fe$_{0.06})_{72.5}$B$_{15}$Si$_{12.5}$ similar to ours, we get 
in Fig. 3, a reasonable agreement with the experimental behaviour, despite a 
faster tapering off of $H_{K}$ at low values of $p$ where we observe 
experimentally a large uncertainty in the value of $H_{K}$, due to a broad 
distribution of anisotropy axis orientation. 

Since our measurement technique provides also the real and imaginary 
components of impedance, we now investigate the frequency behavior of the 
imaginary component of impedance, $X$, for selected values of the DC field, 
Fig. 4, for $p$ = 0.80. For high frequencies ($f >$ 100 MHz), $X$ goes through the 
axis and take negative values; it changes from an inductive to a capacitive 
character. This is a common feature of resonance phenomena. We observe that 
for low DC fields, the plots join in a common point ($f \sim $ 141 MHz, 
$X   \sim  7 \Omega $); for fields larger than the anisotropy field ($H_{K}   
\sim $ 4 Oe), the resonance frequency becomes a function of the field. A 
plot of the imaginary part of impedance as a function of the real component, 
i.e., a Cole-Cole plot, is a more direct evidence of resonance, when the 
locus of the points forms a full circle, as shown in Fig. 5, for the $p$ = 0.8 
sample. A comparison of Cole-Cole plots for the other $p$ values appear in Fig. 
6, for DC field values corresponding to the anisotropy field of each sample. 
The $p$ = 0.32 plot exhibits a deformed circle, probably because the high level 
of mechanical stresses leads to a complex distribution of the anisotropy 
axis. 

High-frequency MI and FMR have led to some 
confusion in the past [16,17]; however, we feel that the differences between 
FMR and MI are now well established. Recent results show [18] that as 
frequency increases (in the 200 MHz -- 6 GHz range for Co-rich amorphous 
ribbons), a divergence in the MI response appears with two maxima in the 
impedance response, corresponding to $\pm H_K$ (the value of which remains
virtually constant at all frequencies), and the FMR response, which becomes 
field dependent, with a Larmor relation that depends 
on the geometry of the sample. FMR experiments in Co-rich amorphous ribbons 
at even higher frequencies (X-band at 9.4 GHz) have exhibited both signals 
clearly resolved [19], the non-resonant low-field absorption (LFA) similar 
to MI at fields lower than 50 Oe, and the expected FMR absorption at 1,682 Oe. 
Therefore our present results show the beginning of this separation. The 
response associated with FMR shows effectively an increase in resonance 
frequency as the field increases as observed in the case $p$ = 0.80.

The relationship between the resonance frequency, $f_{RES}$, and the resonance 
field $H_{RES}$, however, shows a $f ^{3}_{RES}  \sim   H_{RES}$ dependence that does 
not fit the Larmor equation for a cylindrical geometry [20]; 
an FMR experiment at 9.4 GHz in these wires leads to a resonance field of 1,132 Oe,
about two orders of magnitude 
smaller than an extrapolation of the relationship exhibited in Fig. 7. 
In order to understand the behaviour of 
the resonance frequency as a function of the DC field, we follow the work of 
Panina et al. [21] describing the wire as containing an axial core 
magnetization and a circumferential magnetization in an outer shell region 
transverse to the wire axis. From the rotational permeability of the outer 
shell magnetization, one may define a resonance frequency assuming 
negligible Landau-Gilbert damping coefficient:

\begin{equation}
\label{eq1}
f_{RES} = \frac{\gamma }{2\pi }\sqrt {[H_{DC} \sin (\theta + \theta _K ) + H_K 
\cos ^2(\theta ) + 4\pi M_S ][H_{DC} \sin (\theta + \theta _K ) + H_K \cos 
(2\theta )]} 
\end{equation}

where $\gamma $ is the gyromagnetic ratio, $\theta $ is the angle the 
magnetization makes with the circumferential anisotropy axis (CAA) and 
$\theta _{K }$ the angle the CAA makes with a direction perpendicular to 
the wire axis [21]. In principle, $\theta $ is determined from the 
equilibrium condition, however for simplicity we consider the magnetization 
along the CAA (that means we take $\theta $=0) and therefore the simplified 
resonance formula becomes:

\begin{equation}
\label{eq2}
f_{RES} = \frac{\gamma }{2\pi }\sqrt {[H_{DC} \sin (\theta _K ) + H_K + 4\pi M_S 
][H_{DC} \sin (\theta _K ) + H_K ]} 
\end{equation}

Replacing the sin($\theta _{K})$ term by its average value (since the CAA 
fluctuates randomly with respect to the direction perpendicular to the wire 
axis) in the above formula, we are able to make a direct comparison with the 
experimental values we obtain for the resonance frequency versus the DC 
field as displayed in Fig.7. The agreement we obtain is very good given our 
simplifying assumptions and the fact we have no free parameter other than 
the average value of sin($\theta _{K})$. The low field region is a 
crossover region from the domain relaxation to the magnetization rotation in 
the shell whereas in the high field region we ought to observe another 
transition from the magnetization rotation mode to the FMR mode of the axial 
magnetization precession (Fig. 8).

\section{CONCLUSION}

In conclusion, the measurement of the MI response of microwires with a novel 
broadband technique provides a satisfactory view of the interplay between 
different physical phenomena operating in the glass or metal side. In 
addition to the increase of anisotropy field as $p$ decreases, a larger 
distribution of $H_{K}$ is observed for small metal cores. Besides, an 
analysis of the real and imaginary components of impedance has been carried 
out, leading to the observation of a crossover region from the domain 
relaxation to the magnetization rotation in the outer shell. A second 
crossover region to the axial magnetization precession FMR at higher fields 
is observed (see Fig. 8) making the MI a valuable tool to observe and 
identify various modal transitions in these glass-covered microwires. 

Several applications of the present results are possible. One of them is the 
ability to select or tune the physical properties such as a better microwire 
might be produced and suited for a specific application. In particular, the 
presence of a resonant absorption peak at very low fields, that depends on 
$H_{DC}$, allows the possibility of engineering devices that can be designed for 
tunable stopband filtering applications at microwave frequencies [22].

\textbf{Acknowledgments}

The authors acknowledge Prof. M. Vazquez (Spain) for providing the 
microwire samples; R.V. thanks DGAPA-UNAM, Mexico, for partial support 
through grant PAPIIT IN119603-3.

\newpage

\section{FIGURES}

\begin{figure}[!ht]
\begin{center}
\scalebox{0.3}{\includegraphics[angle=0]{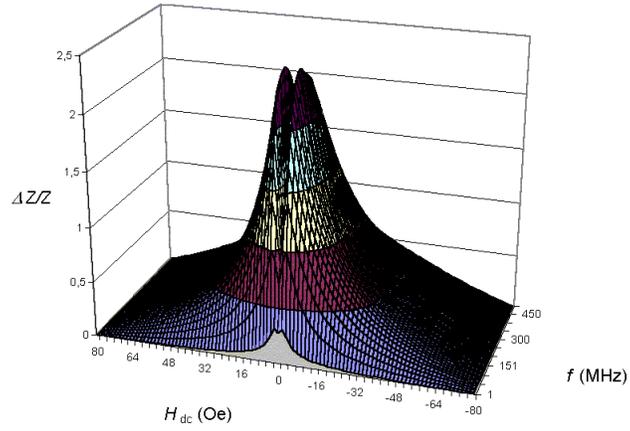}}
\end{center}
\caption{Magnetoimpedance plot for $p$ = 0.8 as a function of axial DC field and 
frequency.}
\label{fig1}
\end{figure}

\begin{figure}[!ht]
\begin{center}
\scalebox{0.3}{\includegraphics[angle=0]{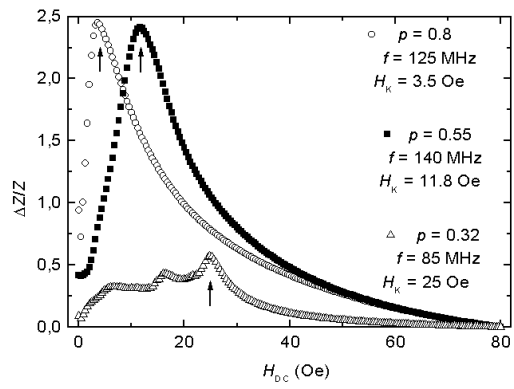}}
\end{center}
\caption{$\Delta Z/Z$ plot as a function of DC field at the selected frequencies 
of the MI maximum for each diameter ratio $p$}
\end{figure}

\begin{figure}[!ht]
\begin{center}
\scalebox{0.8}{\includegraphics[angle=0]{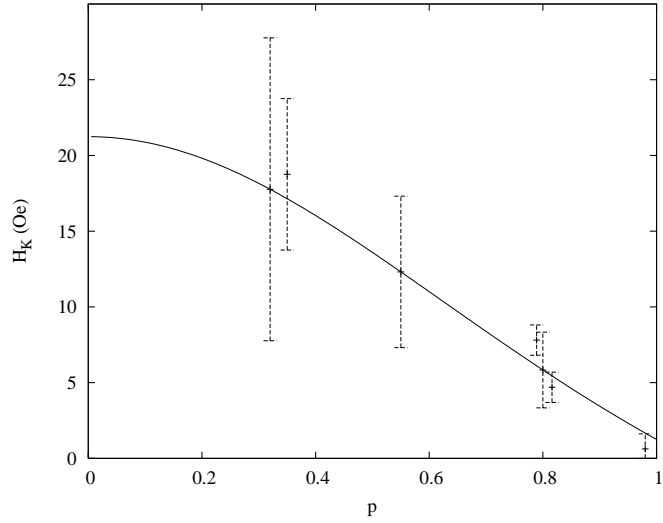}}
\end{center}
\caption{Effect of the diameter ratio $p$ on the anisotropy field $H_{K}$.
Our measured data is for $p=$0.32, 0.55 and 0.8. The $H_{K}$ values corresponding to
$p=$ 0.35, 0.789, 0.816 and  0.98 are adapted from the literature [23].
The values of $\mu _{0}M_{s}$ = 0.8 T, zero stress anisotropy $K_{(\sigma=0)}$~=~40~J/m$^3$, 
and $\lambda_s = -0.1 \times 10^{-6}$ are taken from Ref. [15].
Agreement with the theory improves and uncertainty decreases for larger values of $p$.}
\label{fig3}
\end{figure}

\begin{figure}[!ht]
\begin{center}
\scalebox{0.8}{\includegraphics[angle=0,clip=]{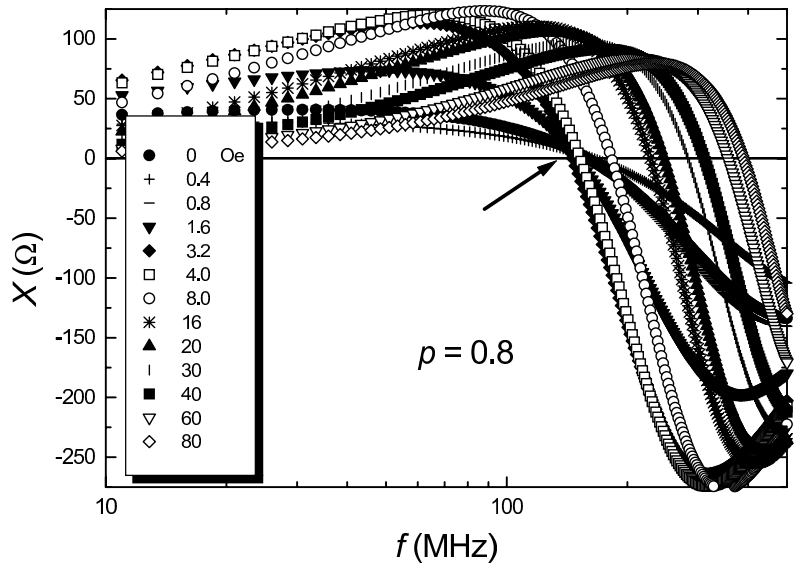}}
\end{center}
\caption{Imaginary part of impedance as a function of frequency, at selected 
values of applied field, for the $p$ = 0.8 sample. The arrow indicates the 
point where plots merge for low values of field.}
\label{fig4}
\end{figure}

\begin{figure}[!ht]
\begin{center}
\scalebox{0.8}{\includegraphics[angle=0]{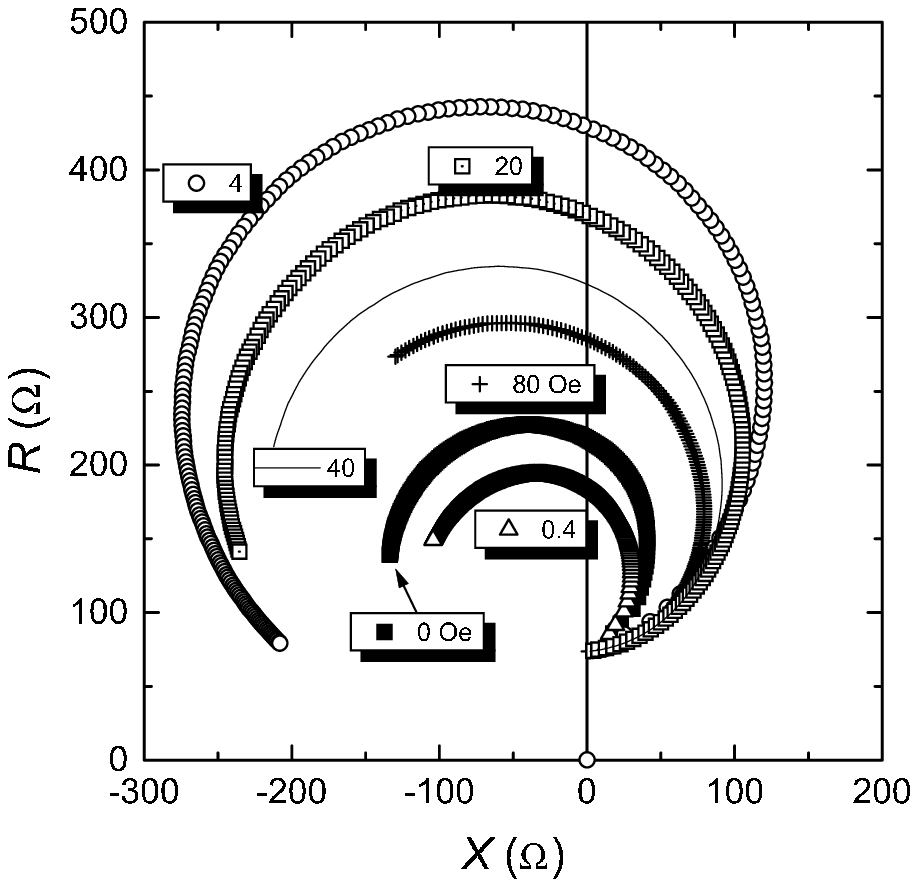}}
\end{center}
\caption{Cole-Cole plot for the $p$ = 0.8 sample, for selected values of the 
applied field.}
\label{fig5}
\end{figure}

\begin{figure}[!ht]
\begin{center}
\scalebox{0.8}{\includegraphics[angle=0,clip=]{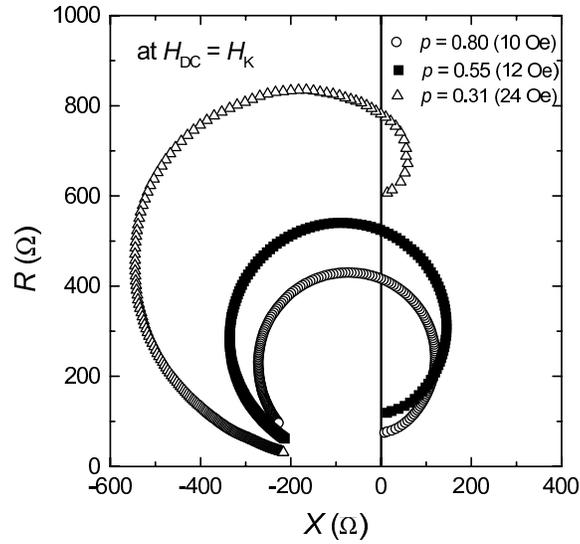}}
\end{center}
\caption{Cole-Cole plots for the three $p$ values, taken at $H_{DC}$ 
corresponding to the anisotropy field value $H_K$.
Note that the $H_K$ are slightly different from those indicated in fig.3.
$H_K$ is taken as 4, 12 and 24 Oe for  $p=$0.8, 0.55 
and 0.32 instead of 3.5, 11.8 and 25 Oe respectively}. 
\label{fig6}
\end{figure}

\begin{figure}[!ht]
\begin{center}
\scalebox{0.6}{\includegraphics[angle=0]{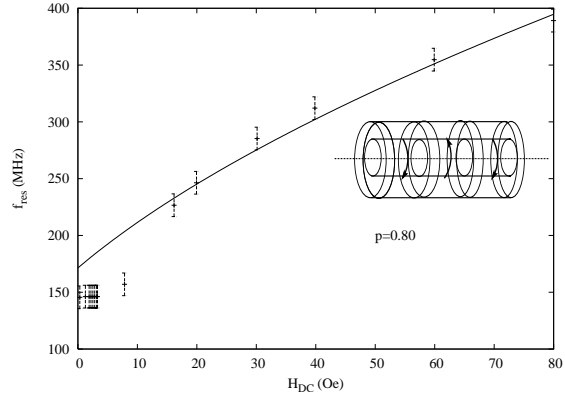}}
\end{center}
\caption{Comparison between the experimentally determined resonance 
frequencies (crosses with errorbars) and the theoretical model (continuous 
line) based on Panina et al's [21] involving rotational motion of the outer 
shell magnetization as shown in the inset. The low field region [0 Oe -10 Oe]
is a crossover region from domain relaxation to magnetization rotation in the shell.}
\label{fig7}
\end{figure}

\begin{figure}[!ht]
\begin{center}
\scalebox{0.4}{\includegraphics[angle=-90]{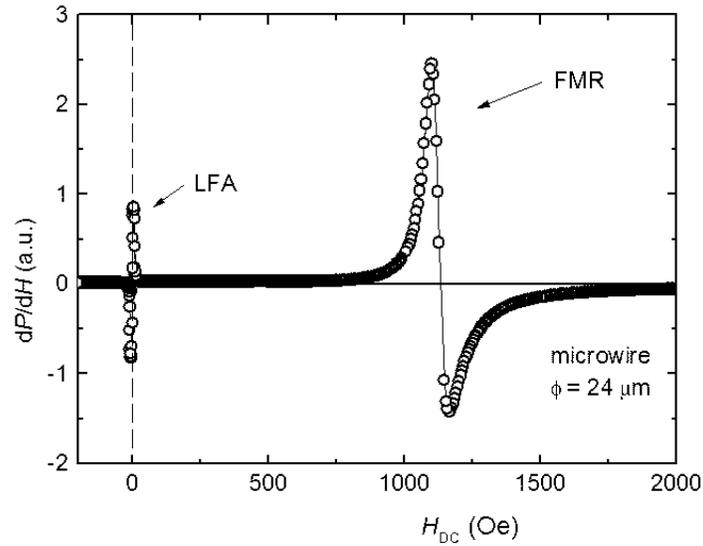}}
\end{center}
\caption{Microwave absorption of $p$ = 0.8 microwire at 9.4 GHz. The FMR 
resonance field is 1,132 Oe. A low-field absorption (LFA) double peak 
appears at $H_{DC}   \sim   H_{K}$ =~4 Oe, where MI was observed at low 
frequencies. The full separation between MI and FMR has therefore taken 
place. }
\label{fig8}
\end{figure}

\end{document}